\documentclass[12pt]{article}
\usepackage{graphicx}
\usepackage{amsmath,amssymb}
\usepackage{bm}
\def\Vec#1{\mbox{\boldmath $#1$}}

\def\itmb{\begin{itemize}}
\def\itme{\end{itemize}}
\def\enmb{\begin{enumerate}}
\def\enme{\end{enumerate}}
\def\eqnb{\begin{equation}}
\def\eqne{\end{equation}}
 
\def\PTP{Prog. Theor. Phys.(Kyoto)}

\def\NPB{{Nucl. Phys.} { B}}

\def\PRL{Phys. Rev. Lett.}
\def\PRD{{Phys. Rev.} D}

\title{Departure from the Standard Model of Meson Decays and Cartan's Supersymmetry}  

\author{Sadataka Furui  \\
 Graduate School of Science and Engineering, Teikyo University\\
2-17-12 Toyosatodai, Utsunomiya, 320-0003 Japan {\thanks
{\textit{E-mail address:} furui@umb.teikyo-u.ac.jp}}
}

\begin{document}
\maketitle
\begin{abstract}%
The experimental decay branching ratios of mesons like $B_s\to \ell\bar \ell$ and $B_d\to \ell\bar \ell$ ($\ell=e$ or $\mu$) do not agree completely with the standard model (SM). 

Cartan's supersymmetry predicts relation of the coupling of vector particles $x_\mu, x_{\mu}'$, ($\mu=1,2,3,4$) to Dirac spinors of large components $\psi, \phi$ and small components $\mathcal C\psi, \mathcal C\phi$.  In the decay of $B_d=\bar b d$, the Cabibbo-Kobayashi-Maskawa(CKM) model suggests that the contribution of $t$ quark dominates, while in the decay of $B_s=\bar b s$, contribution of $c$ quark in the intermediate state is expected to be large, since $s$ and $c$ quark belong to the same CKM sector.  
The relative strength of the $t$ quark and $c$ quark contribution in Cartan's supersymmetric model has more freedom than that of the SM. Together with the problem of enhancement of $B_d\to J/\Psi K_0$ in high $\Delta t$ region, we can understand the problem of branching ratios of $B$ decay into $e\bar e$ and $\mu\bar\mu$, if the Nature follows Cartan's supersymmetry.

\end{abstract}

\newpage
\section{Introduction} 
Recently universality of lepton flavor in the decay of $B$ mesons was questioned by the LHCb group\cite{LHCb13} and conjectures based on the QCD with additional currents were given in \cite{Bobeth13,Bobeth14,AGC14,SSV15}. In \cite{SF15a}, violation of the flavor symmetry due to  admixture of $c$ quark contribution to the $t$ quark contribution was studied, and we want to extend the theory to the decay of $B$ mesons. 

Decay branching ratios of mesons to lepton pairs or a pion and lepton pairs are good testing ground of the standard model (SM). Comparison of experimental decay branching ratios of $B_{s,d}\to \ell\bar\ell$, $K\to \pi\ell\bar\ell$,  $K^+\to \pi^+\nu\bar\nu$ and $K_L\to \pi^0\nu\bar\nu$ with that of SM were discussed in \cite{BB93,BBGT06,BGS10,GH12,IL81,BGNS14,BGHMSS14,CMS14,BGK15,BBR15,BBJ15}.

In the SM the strength of transitions of a quark to different flavor states is defined by the Cabibbo-Kobayashi-Maskawa (CKM) matrices.
For example, the branching ratio of $B_s\to \ell\bar\ell$ depends on the probability of  the coupling of an $s$ quark + emitted vector particle to a $\bar b$ quark + absorbing vector particle which depends on $|V_{cb}|^2\times |V_{tb}^*V_{ts}/V_{cb}|^2$, and the branching ratio of $B_d$ depends on the probability of the coupling of a $d$ quark + emitted vector particle to a $\bar b$ quark + absorbing vector particle which depends on $|V_{tb}^*V_{td}|^2$.

Comparisons of theoretical meson decay matrix elements and experiments were done in various processes, and recently a summary of lattice simulation results was given by the FLAG group \cite{FLAG13}. But the lattice cannot give consistent branching ratios of $B_d$ and $B_s$ mesons together.

Violation of $CP$ symmetry or time reversal symmetry in the decay of $B$ mesons was studied in the difference of
\begin{eqnarray}
B^0\to \ell^-\nu_{\ell}+ X, J/\psi\bar K_L^0 , &{\rm v.s.}\, \bar B^0\to  \ell^+\nu_{\ell}+ X, J/\psi K_L^0  \nonumber\\
B_s\to \ell^-\nu_{\ell}+ X, J/\psi\bar K_L^0 ,&{\rm v.s.}\, \bar B_s\to  \ell^+\nu_{\ell}+ X, J/\psi K_L^0  \nonumber,
\end{eqnarray}
and $CP$ violation was observed in $B^0$ decay but not in $B_s$ decay.
Qualitative differences of $B^0\to\ell\bar\ell$ and $B_s\to\ell\bar\ell$ were also  observed. 

Since complete non leading order corrections in $b\to s\gamma$ and $b\to s\,g$ transitions are not available\cite{BB93}, we apply Cartan's octonion theory to solve the problem of branching ratios of $B$ mesons. 

In \cite{SF12} we studied Cartan's supersymmetry\cite{Cartan66} and studied coupling of  fermions and vector particles which transform under groups $G_{23}, G_{12}, G_{13}, G_{123}$ and $G_{132}$. 

Cartan considered only electromagnetic interactions, but extension to weak interaction is possible\cite{SF15a}. 
 We applied the Cartan's supersymmetric model to $B^0\to K_L J/\Psi$ and found modification of coupling occurs in a certain channel as compared to the SM\cite{SF15b,SF15c}. 

In the section 2, we summarize the status of meson decay branching ratios, and in the section 3 we explain the method of analyzing the meson decay using Cartan's supersymmetry.
Discussion and conclusion are given in section 4.

\section{Deviation of $B$ meson decay branching ratios from the Standard Model}
When one uses results of $K$ meson decay to $\pi\ell\bar\ell$ or $\ell\bar\ell$ states to analyses of $B_d$ meson or $B_s$ decay to $K\ell\bar\ell$ or $\ell\bar\ell$ states, one encounters problems. 

\subsection{$B_d\to K\ell\bar\ell$ and $B_s\to K\ell\bar\ell$}
From the universality of leptons, the strength of $B^+\to K^+\mu\bar\mu$ and that of  $B^+\to K^+ e\bar e$ are expected to be the same, but experimentally
\[
\frac{Br(B^+\to K^+\mu\bar\mu)}{Br(B^+\to K^+ e \bar e)}=0.745^{+0.090}_{-0.074}(stat)\pm 0.036(syst).
\]
is observed by the LHCb group\cite{LHCb13}.  

The inclusive $b\to s\ell\bar\ell$ decay modes were studied by several authors and an extension of Randall and Sundrum model\cite{RS99A,RS99B} to $B\to K\ell\bar\ell$ transition was tried in \cite{BCF14,BCFS14}. 

In the custodially protected Randall Sundrum model (RS$_c$), additional $U(1)$ current in dark sector was proposed to solve $b\to s$ anomalies \cite{AGC14,SSV15}. We do not assume this current, but adopt the vertex with $1-\gamma_5$ factor for the weak interaction, and consider two triangle diagrams in the coupling.
\begin{figure}[htb]
\begin{minipage}[b]{0.47\linewidth}
\begin{center}
\includegraphics[width=5cm,angle=0,clip]{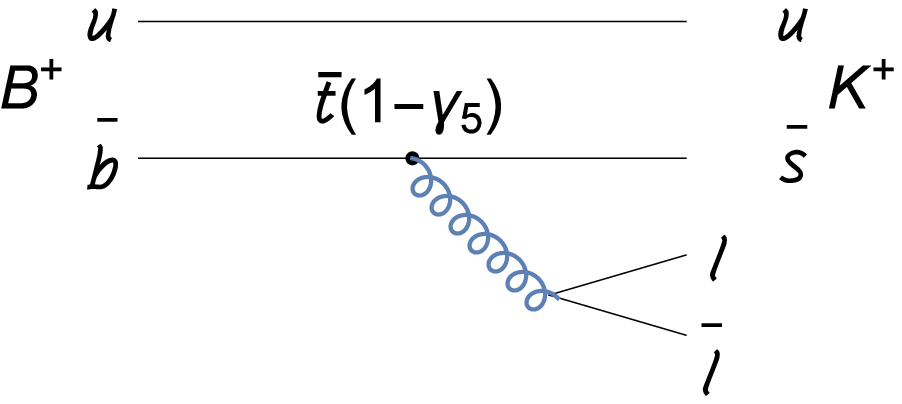} 
\end{center}
\end{minipage}
\hfill
\begin{minipage}[b]{0.47\linewidth}
\begin{center}
\includegraphics[width=5cm,angle=0,clip]{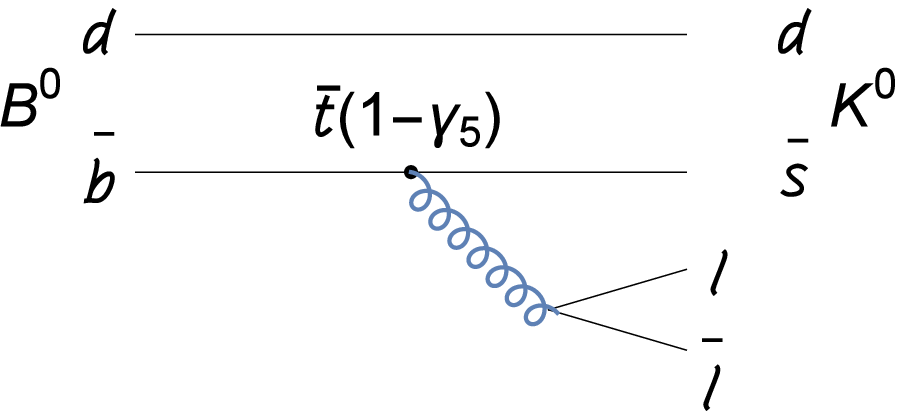}   
\end{center}
\end{minipage}
\caption{The vertex of $B^+\to K^+\ell\bar\ell$ (left) and ${B^0}\to {K^0}\ell\bar\ell$ (right).}
\label{B0K0LLTr}
\end{figure}

The $b\to s\ell\bar\ell$ vertex requires renormalization of penguin diagrams.
\begin{figure}[htb]
\begin{minipage}[b]{0.47\linewidth}
\begin{center}
\includegraphics[width=5cm,angle=0,clip]{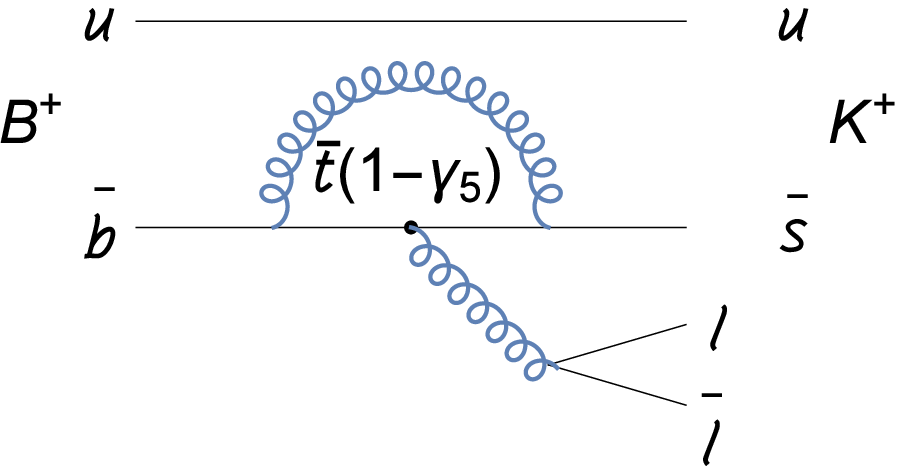} 
\end{center}
\end{minipage}
\hfill
\begin{minipage}[b]{0.47\linewidth}
\begin{center}
\includegraphics[width=5cm,angle=0,clip]{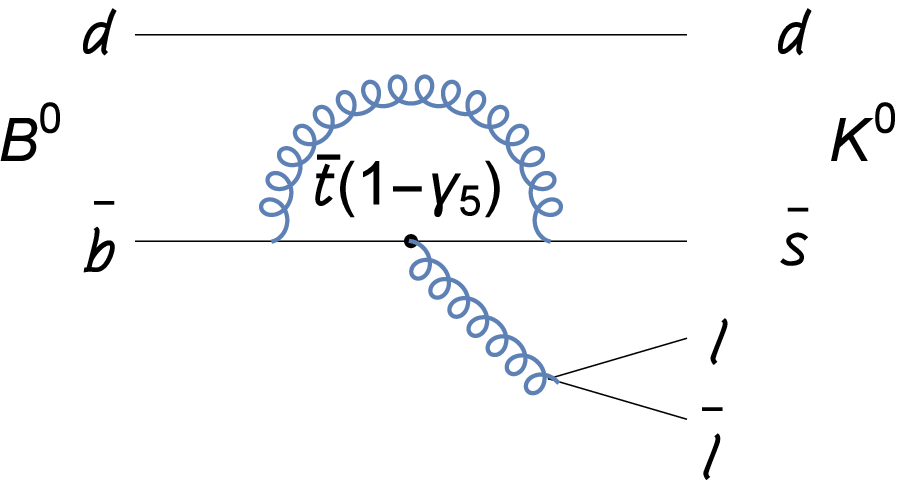}   
\end{center}
\end{minipage}
\caption{The penguin diagram of $B^+\to K^0\ell\bar\ell$ (left)
and  ${B^0}\to {K^0}\ell\bar\ell$ (right).}
\label{B0K0LLPn}
\end{figure}

The lepton pairs can be $e\bar e$ which transform mainly to $\gamma$ or $\mu\bar \mu$ which transform mainly to $Z$. The CKM model implies higher order effect of coupling of $\gamma$ to $b$ quark is dominated by the loop of $t$ quarks, but difficulty of the CKM model to explain the decay of $B\to K+X$ transitions implies contamination of the loop of $b-c-s$ quarks in addition to $b-t-s$ quarks, since $s$ and $c$ quark belong to the same CKM sector.  
\begin{figure}[htb]
\begin{minipage}[b]{0.47\linewidth}
\begin{center}
\includegraphics[width=5cm,angle=0,clip]{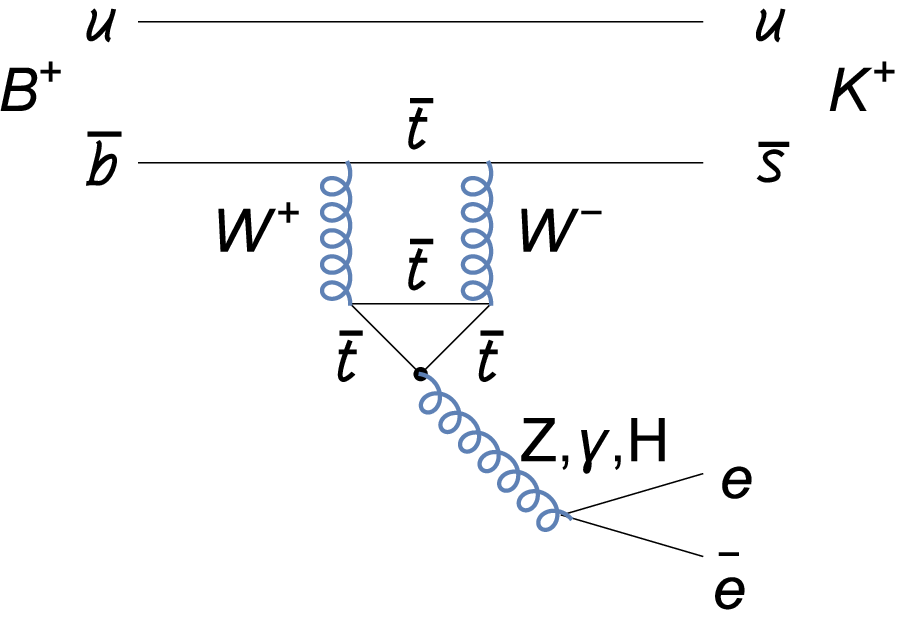} 
\end{center}
\end{minipage}
\hfill
\begin{minipage}[b]{0.47\linewidth}
\begin{center}
\includegraphics[width=5cm,angle=0,clip]{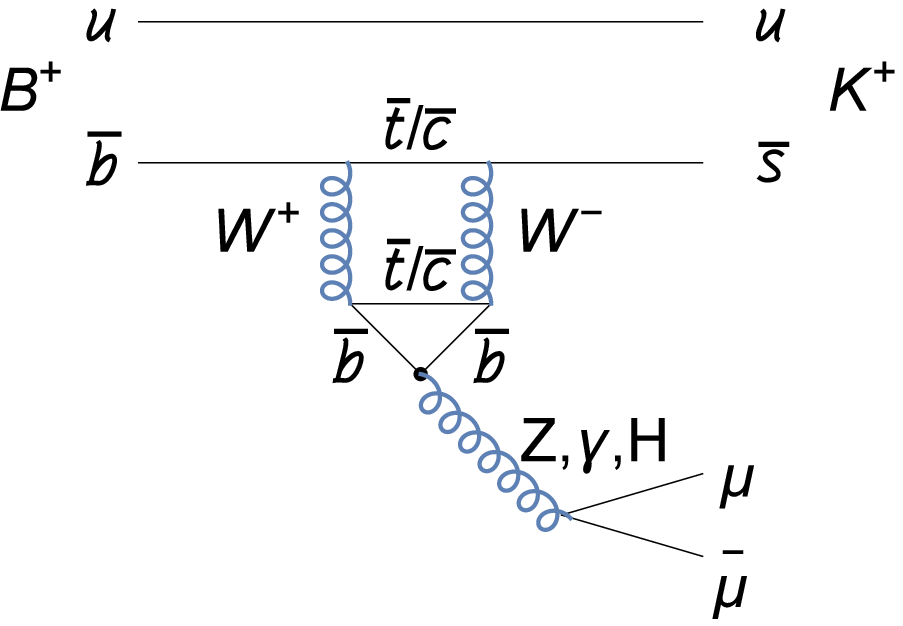}  
\end{center}
\end{minipage}
\caption{The Feynman diagrams including two loops of $B^+\to K^+e\bar e$ (left),
and ${B^+}\to {K^+}\mu\bar\mu$ (right).}
\label{BsMuMuW}
\end{figure}

\newpage
\subsection{$B_s\to \ell\bar\ell$ and $B_d\to\ell\bar\ell$}

\begin{figure}[htb]
\begin{minipage}[b]{0.47\linewidth}
\begin{center}
\includegraphics[width=6cm,angle=0,clip] {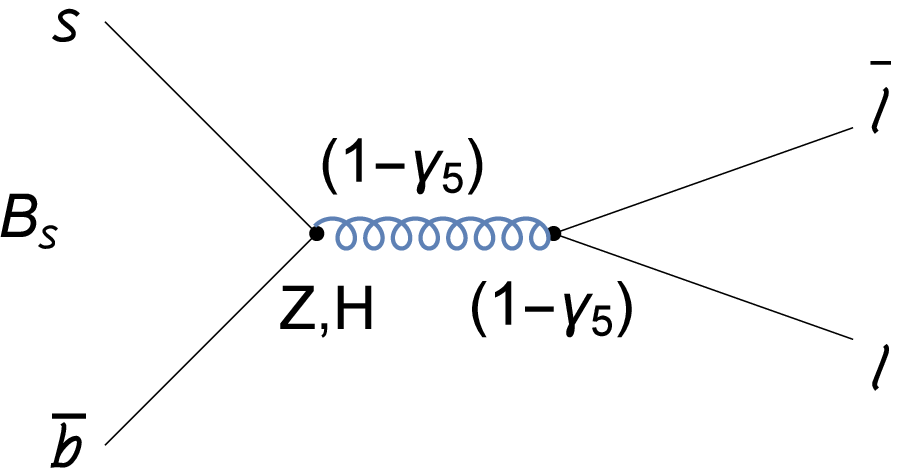}%
\end{center}
\end{minipage}
\hfill
\begin{minipage}[b]{0.47\linewidth}
\begin{center}
\includegraphics[width=6cm,angle=0,clip]{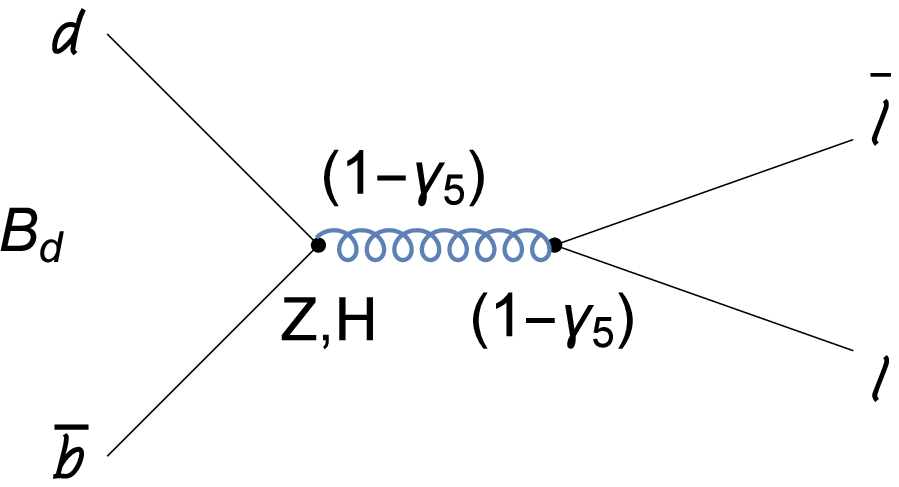}  
\end{center}
\end{minipage}
\caption{The electro-weak interaction quark diagrams of  $B_s\to\ell\bar\ell$ (left), and $B_d\to\ell\bar\ell$ (right).} 
\label{bbarsd}
\end{figure}

The decay of $B_{s,d}\to\ell^+\ell^-$ ($\ell=e,\mu$ and $\tau$) in the standard model contains two $\gamma_5$ verices and the intermediate state is $Z$ or $H$, as shown in Fig.\ref{bbarsd}.

They defined the Lagrangian for $B_q\to\ell\bar\ell$ as
\[
\mathcal L_{weak}=N\, C_A(\mu_b)(\bar b\gamma_a \gamma_5 q)(\bar\ell \gamma^a \gamma_5 \ell)+O(\alpha_{em})
\]
and averaged time-integrated branching ratios proportional to
\begin{eqnarray}
\bar B_{s\ell}&\propto&|N|^2 |C_A(\mu_b)|^2 \tau_H^s\nonumber\\
&\propto&f_{B_s}^2 |V_{cb}|^2 (|V_{tb}^* V_{ts}/V_{cb}|)^2 \tau_H^s,
\end{eqnarray}
where $\tau_H^s=1/\Gamma_H^s$ is the lifetime of the heavior mass eigenstate of $B$\cite{HFAG12}, $V_{tb}, V_{ts}$ and $V_{cb}$ are CKM matrix elements, was calculated for $B_s\to\mu^+\mu^-$, and 
\[
\bar B_{d\ell}\propto f_{B_d}^2|V_{tb}^* V_{td}|^2 \tau_d^{av}
\]
where $\tau_d^{av}=2/(\Gamma_H^d+\Gamma_L^d)$ is the average of the lifetime of the lighter mass and heavior mass eigenstate of $B$,  $V_{tb}$ and $V_{td}$ are CKM matrix elements, was calculated for $B_d\to\mu^+\mu^-$. 

 In the SM, the decay branching ratio of $B_d$ and $B_s$ yields direct information on $V_{bt}, V_{st}$ and $V_{bc}$,  but uncertainty in the CKM matrix element $V_{bc}$ was not seriously taken into account.

Blanke et al.\cite{BBR15} presented the review of  $K_L\to\ell\bar\ell$, $B\to K\ell\bar\ell$ and $B\to \ell\bar\ell$. According to their work experimental world average branching ratio
\[
{\bar B}( B_s\to\mu^-\mu^+)=(2.9\pm 0.7)\times 10^{-9}
\]
is slightly smaller than the theoretical value $(3.65\pm 0.23)\times 10^{-9}$\cite{BGHMSS14}.

In the case of $B_d\to \mu^-\mu^+$, there are larger experimental uncertainty\cite{CMS14},
however the experimental world average branching ratio
\[
{\bar B}(B_d\to\mu^-\mu^+)=(3.6_{-1.4}^{+1.6})\times 10^{-10}
\]
is much larger than the theoretical value $(1.06\pm 0.09)\times 10^{-10}$\cite{BGHMSS14}.

A two loop calculation of branching ratio including the diagram of the penguin diagram was done by \cite{GH12} for the $K_L\to\mu^+\mu^-$. If a neutrino is a Majorana neutrino and $\nu$ and $\bar\nu$ are indistinguishable, one can consider the decay through the box diagram\cite{IL81,GH12}. But there is no clear experimental results that suggests $\nu|_{Majorana}=\bar\nu|_{Majorana}$. If neutrino is massive, it is difficult to identify a neutrino and an antineutrino. 

We expect the decay of $K_L\to\mu^+\mu^-$ and $B\to\mu^+\mu^-$ are not dominated by Fig.\ref{bbarsd}, we study $c$ quark contribution is important. In the case of $K_L\to\mu^+\mu^-$ we consider the $q-\bar q-W$ triangle diagram with a penguin loop studied as \cite{GH12} and the $W^+-W^- -q$ triangle diagram.  We do not assume that the penguin loop is given by a gluon but by a $Z$ boson. A gluon has unphysical components and it can be replaced by a gauge potential and ghost fields\cite{BeRi14}.

Since the $s$ quark and the $c$ quark are in the same CKM sector, we expect in the $q-\bar q-W$ triangle, some contribution of $c\bar c$ quark to the usual $t\bar t$ quark contribution. Since a $c$ quark is heavior than a $s$ quark, the contribution is less important than $B_d\to\ell\bar\ell$ or $B_s\to\ell\bar\ell$.

In \cite{SF15b} we analyzed $B_d\to K_L J/\Psi$ decay process using Cartan's supersymmetry and compared the results with SM using CKM matrix elements. We perform the similar analysis for $K$ meson and $B$ meson decays into lepton pairs.

In the previous analysis of $B_d\to K_L J/\Psi$, a vector particle was emitted from a quark or an anti-quark in a meson and transformed to $J/\Psi$, but in the present case we consider processes in which a vector particle transform to $\ell\bar\ell$. 

The coupling of the vector particle which was the source of $J/\Psi$ was pseudo scalar type
$
{^t\bar \psi} \gamma_\mu\gamma_5\psi.
$

The agreement of branching ratios of  meson decays in theoretical models and the SM is not sufficient. One could imagine new physics, in which heavy mesons $X^+X^0$ instead of $W^+ Z$ contribute\cite{CMS14}.  However, it may be appropriate to study supersymmetry of quarks or leptons and vector particles more in detail.

\section{Cartan's supersymmetry and meson decays} 

In the SM, the simplest amplitudes of $B\to\ell\bar\ell$ contains two $\gamma_5$ vertices, since electromagnetic decay is impossible.   
In the Cartan's supersymmetric model, $W^+, W^-$ and $Z$ bosons are replaced by vector particles $x_1,x_2, x_3, x_4, x_1', x_2', x_3'$ and $x_4'$, and quarks and leptons are replaced by Dirac spinors $(\psi, \mathcal C\psi)$ and $(\phi,\mathcal C\phi)$. Incorporation of $\phi$ quarks allows decay amplitudes containing only one $\gamma_5$ vertex from a part of 
${^t\bar \phi}\gamma_\mu(1-\gamma_5)\psi$.

\subsection{$K_L\to\ell\bar\ell$ and $B_{s,d}\to\ell\bar\ell$}
\begin{figure}[htb]
\begin{minipage}[b]{0.47\linewidth}
\begin{center}
\includegraphics[width=6cm,angle=0,clip] {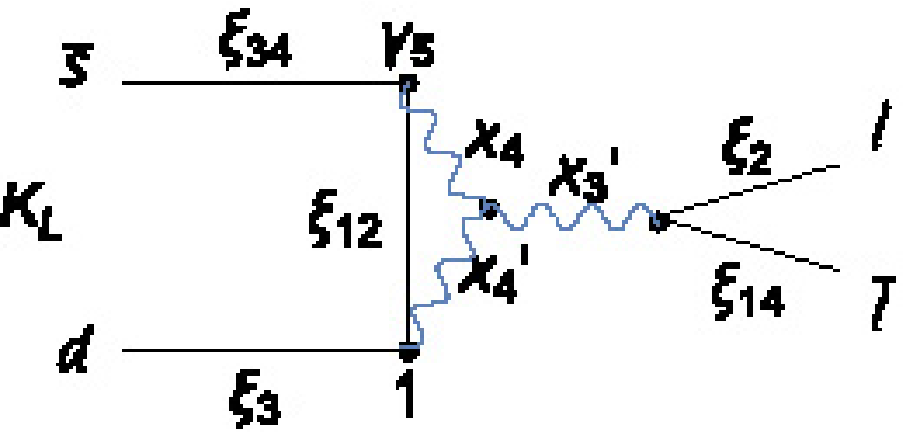}%
\end{center}
\end{minipage}
\hfill
\begin{minipage}[b]{0.47\linewidth}
\begin{center}
\includegraphics[width=6cm,angle=0,clip]{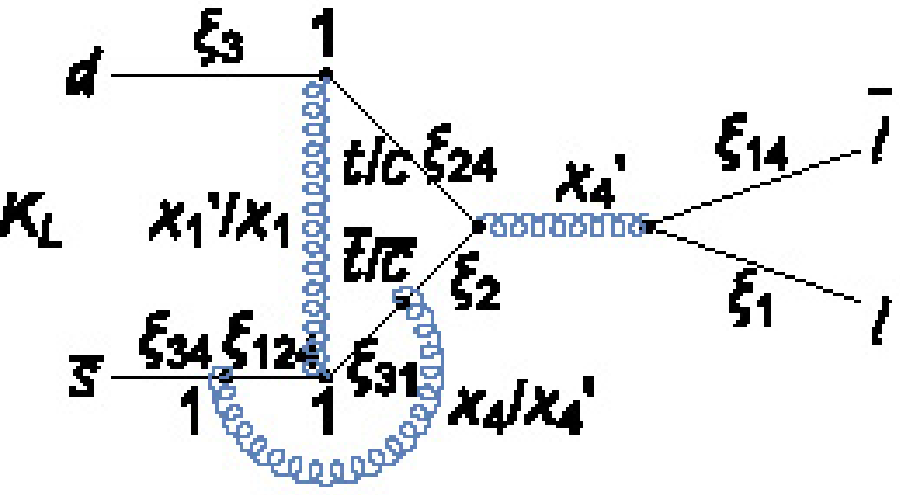}  
\end{center}
\end{minipage}
\begin{minipage}[b]{0.47\linewidth}  
\begin{center}
\includegraphics[width=6cm,angle=0,clip]{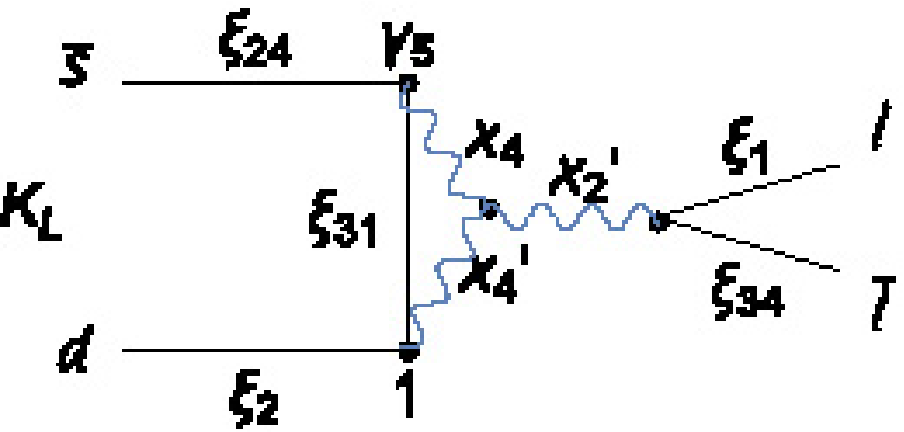}%
\end{center}
\end{minipage}
\hfill
\begin{minipage}[b]{0.47\linewidth}
\begin{center}
\includegraphics[width=6cm,angle=0,clip]{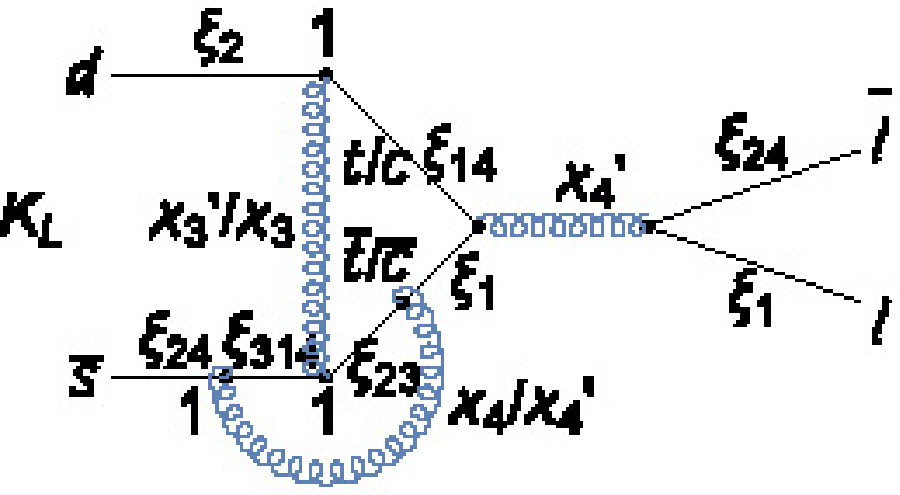}%
\end{center}
\end{minipage}
\begin{minipage}[b]{0.47\linewidth}
\begin{center}
\includegraphics[width=6cm,angle=0,clip]{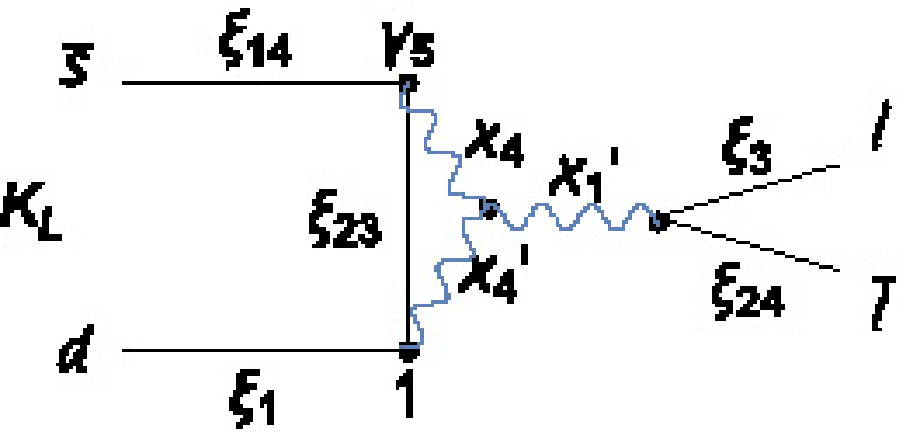}%
\end{center}
\end{minipage}
\hfill
\begin{minipage}[b]{0.47\linewidth}
\begin{center}
\includegraphics[width=6cm,angle=0,clip]{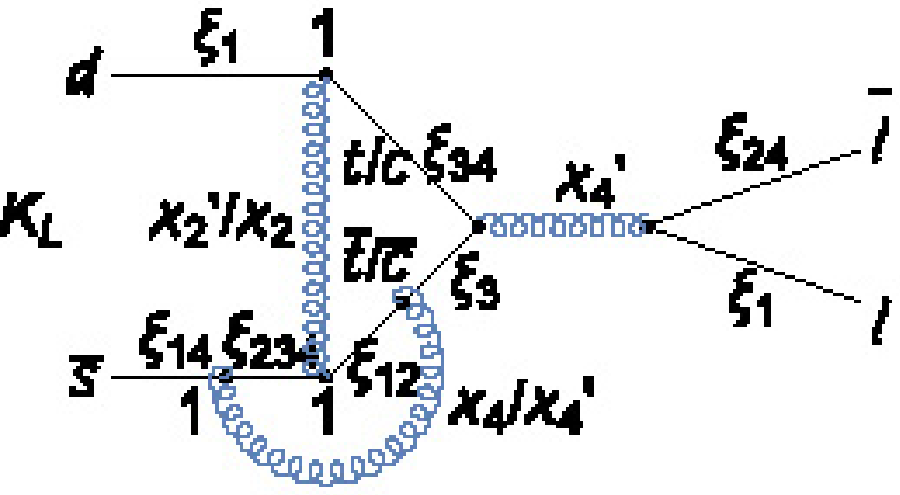}       %
\end{center}
\end{minipage}
\begin{minipage}[b]{0.47\linewidth}
\begin{center}
\includegraphics[width=6cm,angle=0,clip] {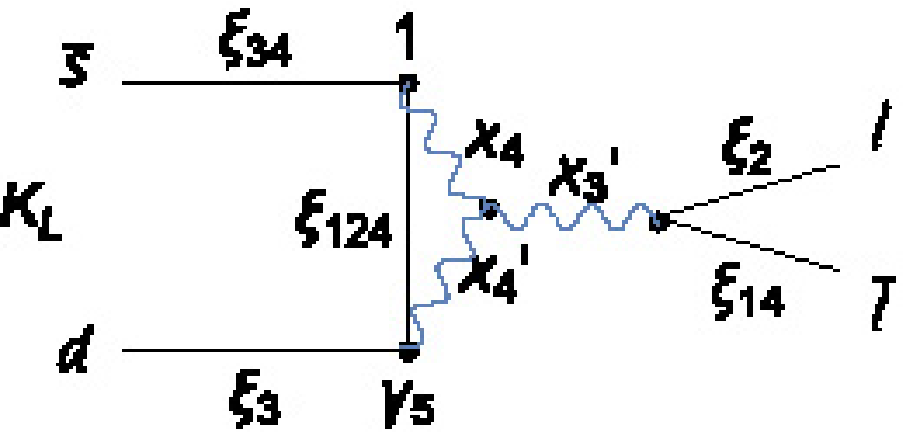}%
\end{center}
\end{minipage}
\hfill
\begin{minipage}[b]{0.47\linewidth}
\begin{center}
\includegraphics[width=6cm,angle=0,clip]{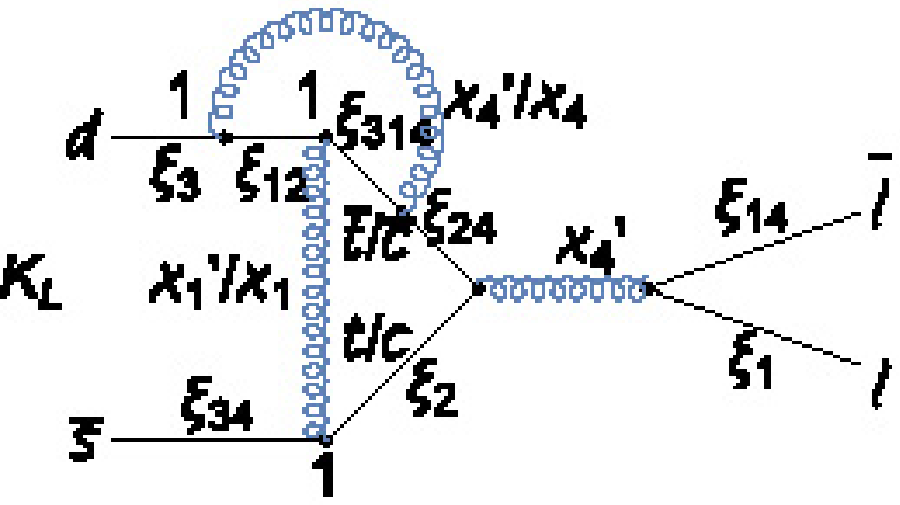}  
\end{center}
\end{minipage}
\begin{minipage}[b]{0.47\linewidth}  
\begin{center}
\includegraphics[width=6cm,angle=0,clip]{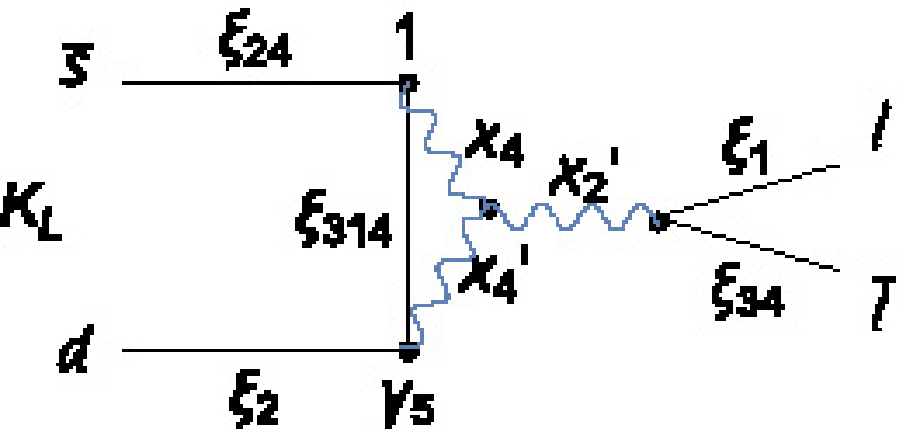}%
\end{center}
\end{minipage}
\hfill
\begin{minipage}[b]{0.47\linewidth}
\begin{center}
\includegraphics[width=6cm,angle=0,clip] {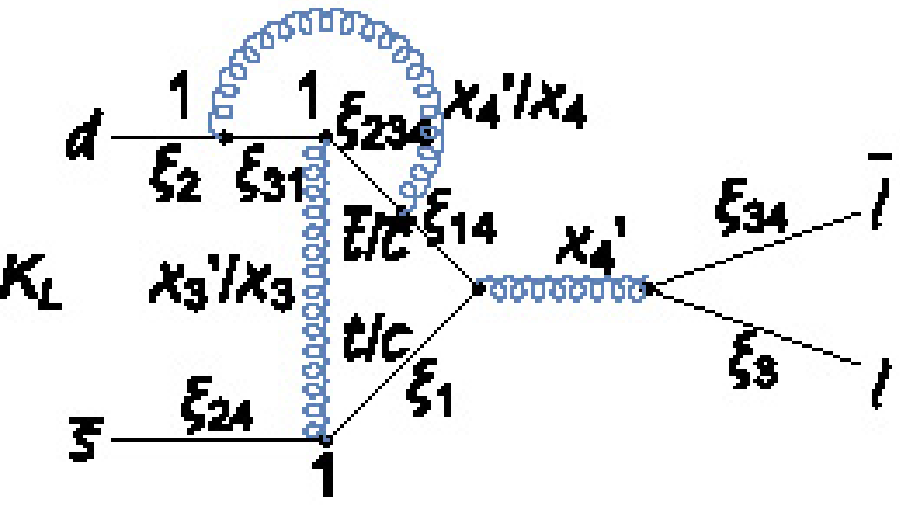}%
\end{center}
\end{minipage}
\begin{minipage}[b]{0.47\linewidth}
\begin{center}
\includegraphics[width=6cm,angle=0,clip]{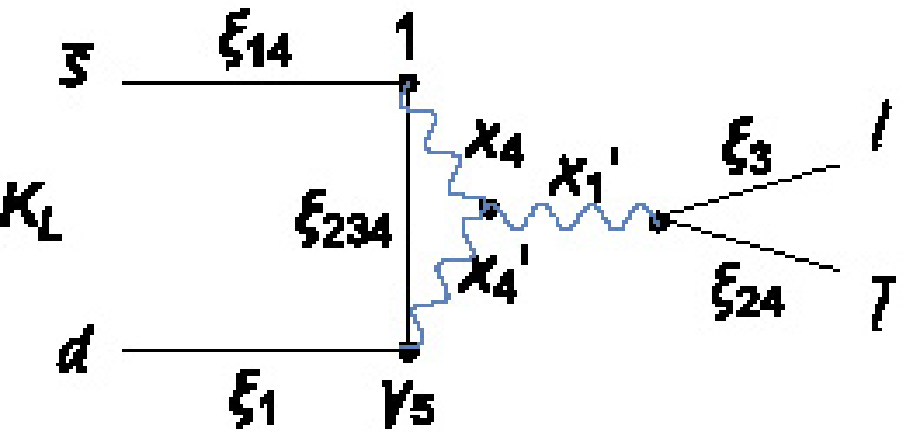}%
\end{center}
\end{minipage}
\hfill
\begin{minipage}[b]{0.47\linewidth}
\begin{center}
\includegraphics[width=6cm,angle=0,clip]{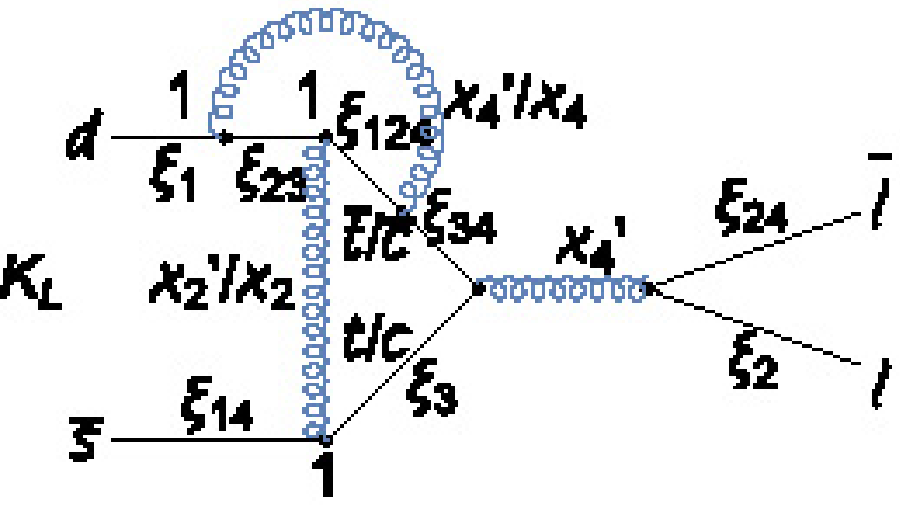}       %
\end{center}
\end{minipage}
\caption{$K_L\to \ell\bar\ell$ quark diagrams containing $x_4 {x_4}' {x_k}'$ ($k=1,2,3$) vertices with different $\gamma_5$ positions (left), and $q-\bar q-W$ triangle diagrams with different penguin loop positions (right).} 
\label{sbardLLa}
\end{figure}

Simple diagrams of  $K_L\to \ell\bar\ell$ containing the vertex corresponding to that of $W^+W^- Z$ are shown in Fig.\ref{sbardLLa}. 

In our diagram of $K_L\to\ell\bar\ell$ shown in the left side of Fig.\ref{sbardLLa}, the symbol $x_3'$ corresponds to $Z$ boson with the momentum direction $x_3'$, the symbol $x_4$ corresponds to $W^+$ with the momentum direction $x_4$, and the symbol $x_4'$ corresponds to $W^-$ with the momentum direction $x_4'$.
The vector particle $x_4$ couple with $\xi_{12}$ of $\mathcal C\phi$ and create $\xi_{34}$ of anti-quark $\phi$ and the vector particle $x_4'$ is emitted from $\xi_{3}$ of $\psi$ and create the $\xi_{12}$. 

In order to make coupling of a quark to $W^+$ and $W^-$ bosons, it is necessary to include a $\gamma_5$ vertex which couples $\phi$ and $\mathcal C\psi$ or $\psi$ and $\mathcal C\phi$. A quark emitted as $\mathcal C\phi=\xi_{12}$ can be absorbed as $\mathcal C\psi=\xi_{124}$ at the $\gamma_5$ vertex.

We extend the Cartan's supersymmetric model to allow a spinor
\[
\tilde \psi=\xi_1\Vec i+\xi_2\Vec j+\xi_3\Vec k
\]
and an antispinor
\[
\tilde \phi=\xi_{14}\Vec i+\xi_{24}\Vec j+\xi_{34}\Vec k
\]
annihilate to a scalar particle $Z=x_4'$. Including the vertex $\xi_i\xi_{i4} Z$ vertices, we
can consider penguin diagrams of \cite{GH12}, which are shown in the right hand side of  
Fig.\ref{sbardLLa}. In the diagrams there are triangle diagrams of $q(t/c)-\bar q(\bar t/\bar c)-W$ boson $(x_i'$) and a loop of $Z$ boson $(x_4'/x_4)$. Due to this loop, coupling of a quark to the $Z$ boson($x_4'$) becomes possible, without including a $\gamma_5$ vertex. They appear as a correction from weak interactions.

In our model, we allow transformation of $x_i$ to $x_{i'}$ during propagations, and since $K_L$ consists of $\bar s d$, and $\bar s$ quark and $\bar c$ quark are in the same sector of CKM model, the $q(\bar q)$ on the triangle is assumed to be $t/c (\bar t/\bar c)$.
In other words, we take into account the correction to the $t$ quark loop dominance in the SM.  

In the calculation of the right hand side of Fig.\ref{sbardLLa}, we fix the $d$ quark and the $\bar s$ quark, both large components and exchange a vector particle between $d$ and $\bar s$. In the first diagram, the $d$ quark $\xi_3$ becomes after emitting a vector particle $x_{1'}$ becomes a $\bar t$ quark or a $\bar c$ quark $\xi_{24}$, and after emitting a particle $x_{4'}$ corresponding to a $Z$ boson becomes a quark $\xi_2$ and after emitting $x_4'$, becomes an anti-quark $\xi_{31}$. The $\bar s$ quark $\xi_{34}$ emits a $x_4$ and becomes a quark $\xi_{124}$. The $\mathcal C\psi\mathcal C\phi$ state of $\xi_{124}\xi_{31}$ has an overlap with the vector particle $x_1$ and absorbs the previously emitted $x_1'$. The $Z$boson $x_4'$ becomes an $\ell\bar \ell$ pair.
By admitting an overlap of $x_1x_4'$ and $x_1'x_4$ state, $d\bar s\to \ell\bar\ell$ transition becomes possible.  The contribution of $c$ or $\bar c$ quark is expected to be not so important since $m_c>m_s$ and virtual transition to $t$ whose mass $m_t>m_H$ is not expected to be affected.

In the case of $B_d$ we replace the $\bar s$ quark to $\bar b$ quark. In the case of $B_s$ we replace the $d$ quark in $B_d$ to an $s$ quark and $t/\bar t$ quark to $c/\bar c$ quark, since a $c$ quark is in the same CKM sector as an $s$ quark, and $m_c<m_b$, we need to incorporate the effect of a $c$ quark. 

\begin{figure}[htb]
\begin{minipage}[b]{0.47\linewidth}
\begin{center}
\includegraphics[width=6cm,angle=0,clip]{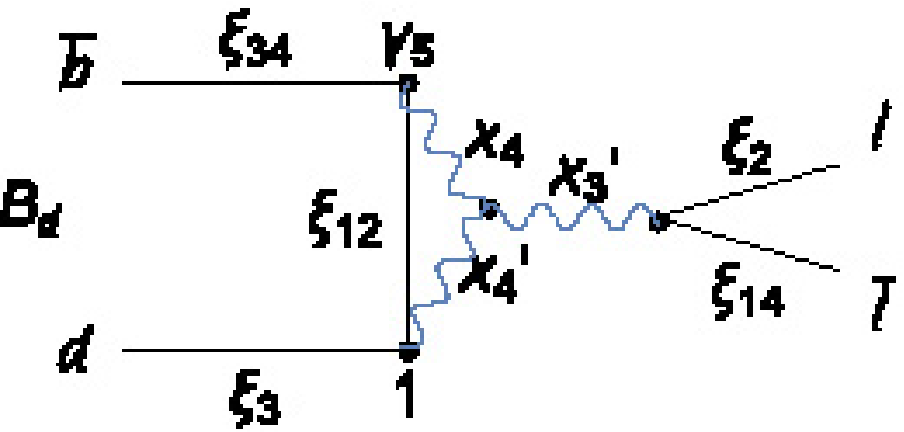}%
\end{center}
\end{minipage}
\hfill
\begin{minipage}[b]{0.47\linewidth}
\begin{center}
\includegraphics[width=6cm,angle=0,clip]{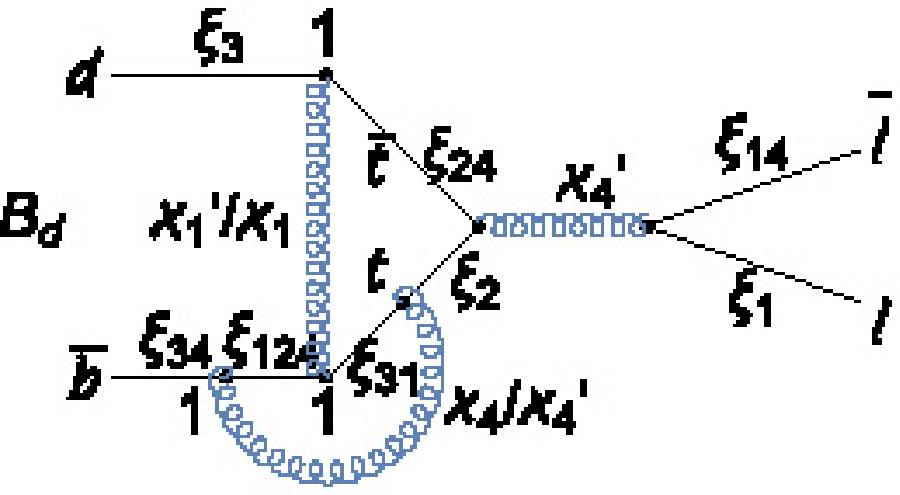}%
\end{center}
\end{minipage}
\begin{minipage}[b]{0.47\linewidth}
\begin{center}
\includegraphics[width=6cm,angle=0,clip]{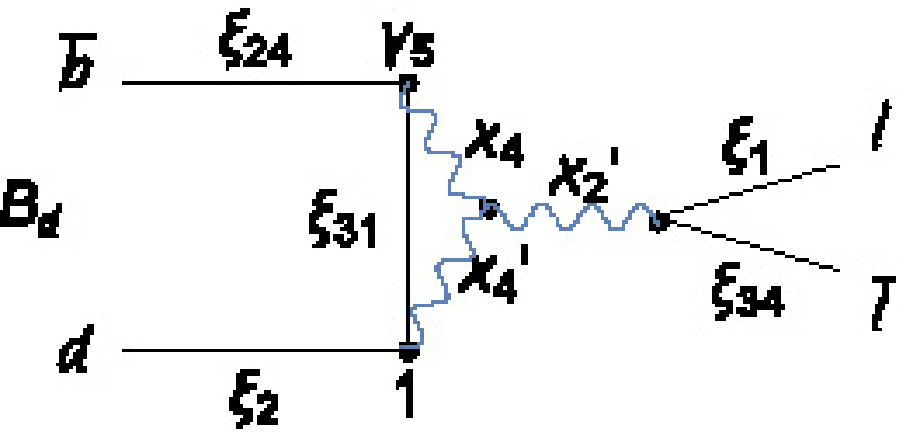}%
\end{center}
\end{minipage}
\hfill
\begin{minipage}[b]{0.47\linewidth}
\begin{center}
\includegraphics[width=6cm,angle=0,clip]{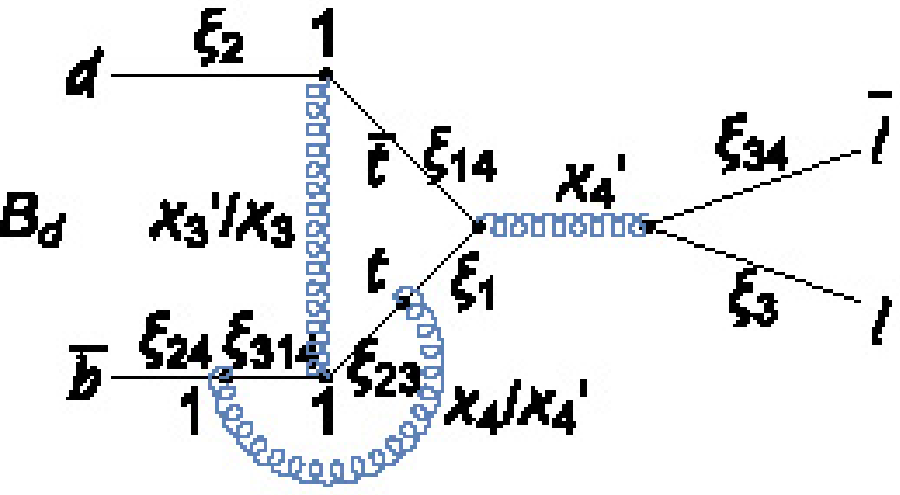}%
\end{center}
\end{minipage}
\begin{minipage}[b]{0.47\linewidth}
\begin{center}
\includegraphics[width=6cm,angle=0,clip]{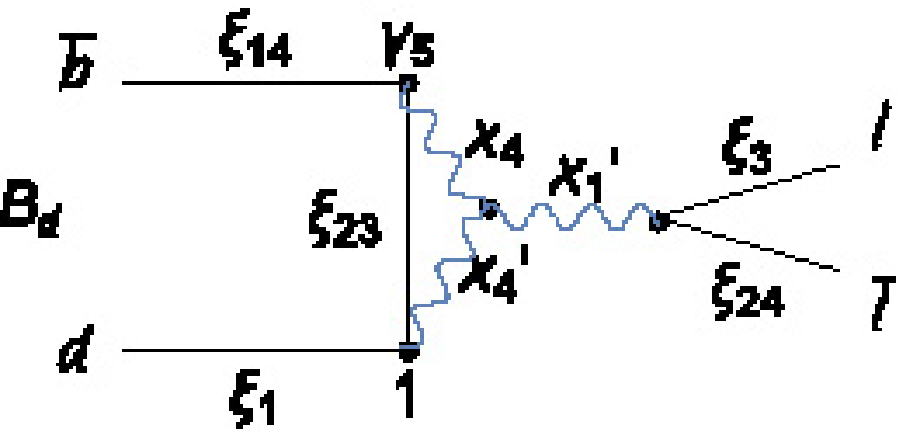}%
\end{center}
\end{minipage}
\hfill
\begin{minipage}[b]{0.47\linewidth}
\begin{center}
\includegraphics[width=6cm,angle=0,clip]{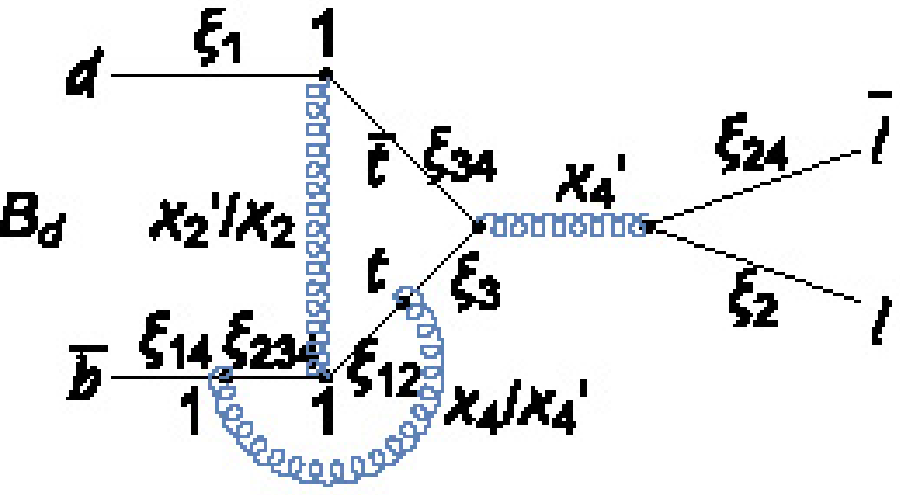}%
\end{center}
\end{minipage}
%
\begin{minipage}[b]{0.47\linewidth}
\begin{center}
\includegraphics[width=6cm,angle=0,clip]{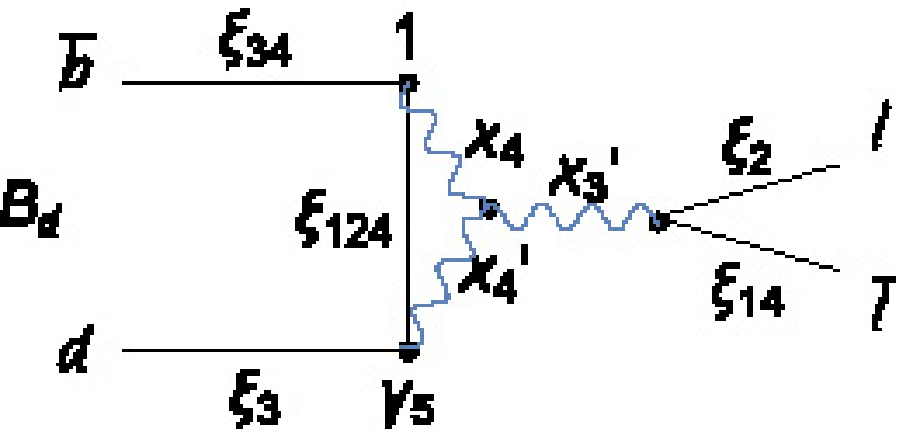}%
\end{center}
\end{minipage}
\hfill
\begin{minipage}[b]{0.47\linewidth}
\begin{center}
\includegraphics[width=6cm,angle=0,clip]{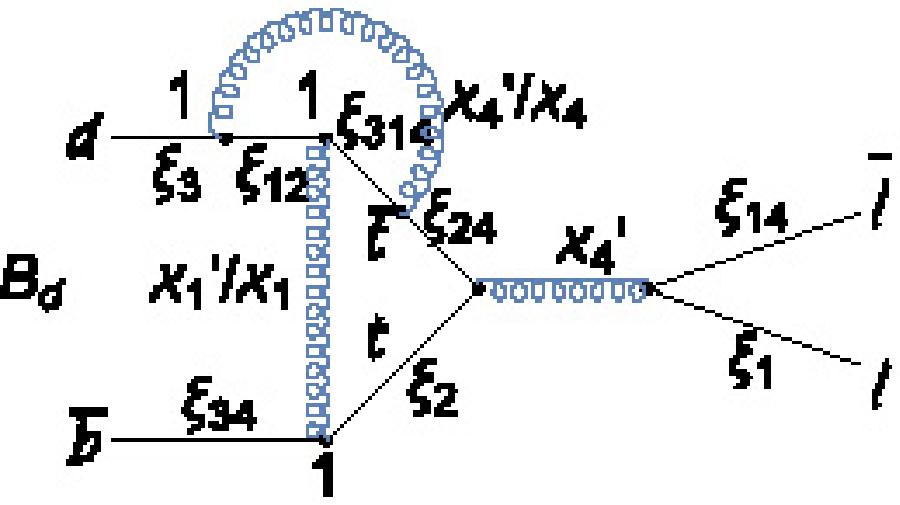}%
\end{center}
\end{minipage}
\begin{minipage}[b]{0.47\linewidth}
\begin{center}
\includegraphics[width=6cm,angle=0,clip]{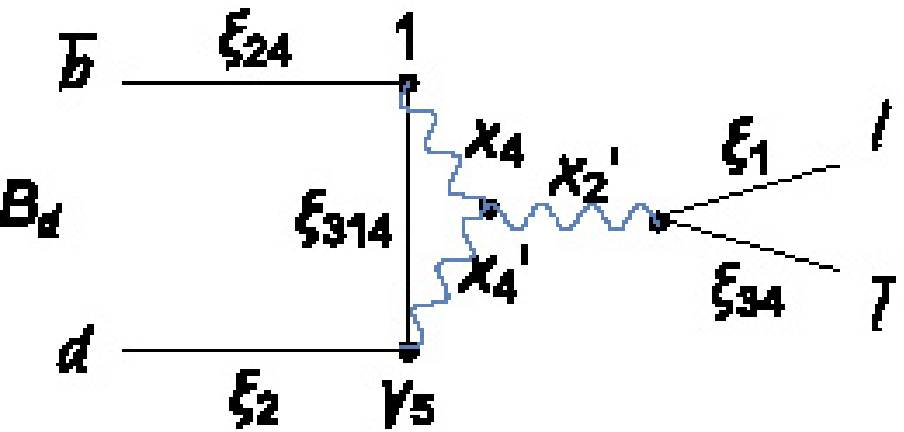}%
\end{center}
\end{minipage}
\hfill
\begin{minipage}[b]{0.47\linewidth}
\begin{center}
\includegraphics[width=6cm,angle=0,clip]{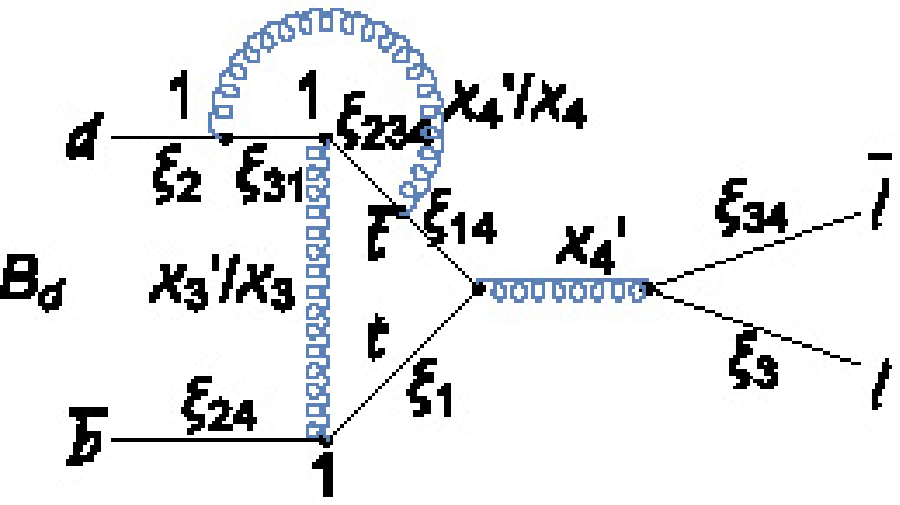}%
\end{center}
\end{minipage}
\begin{minipage}[b]{0.47\linewidth}
\begin{center}
\includegraphics[width=6cm,angle=0,clip]{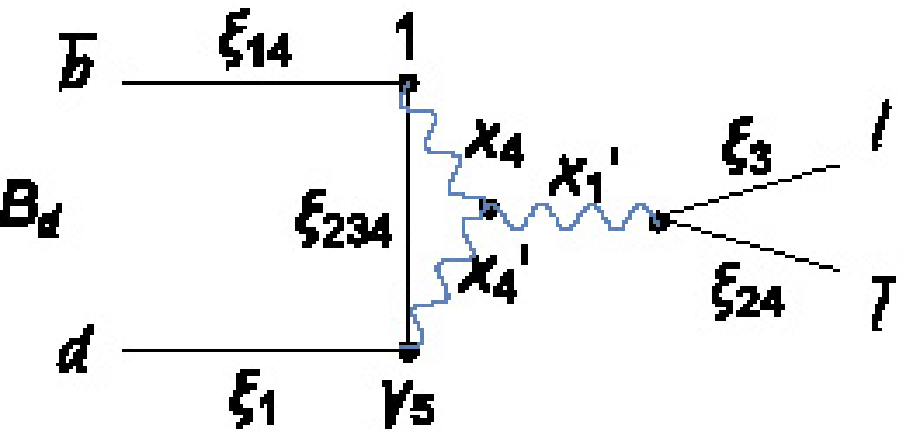}%
\end{center}
\end{minipage}
\hfill
\begin{minipage}[b]{0.47\linewidth}
\begin{center}
\includegraphics[width=6cm,angle=0,clip]{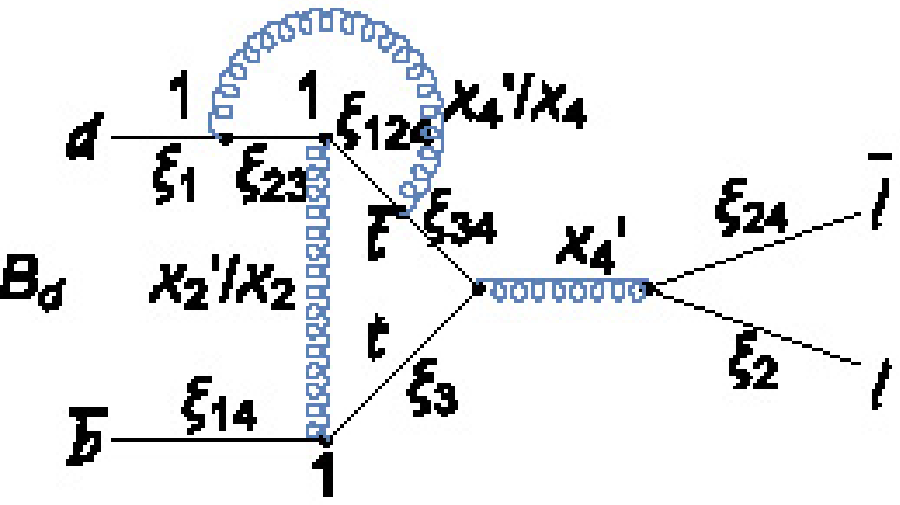}%
\end{center}
\end{minipage}
\caption{$B_d\to \ell\bar\ell$ quark diagrams containing $x_4 {x_4}' {x_k}'$ ($k=1,2,3$) vertices with different $\gamma_5$ positions (left), and $q-\bar q-W$ triangle diagrams with different penguin loop positions (right).} 
\label{bbardLLa}
\end{figure}

\begin{figure}[htb]
\begin{minipage}[b]{0.47\linewidth}
\begin{center}
\includegraphics[width=6cm,angle=0,clip]{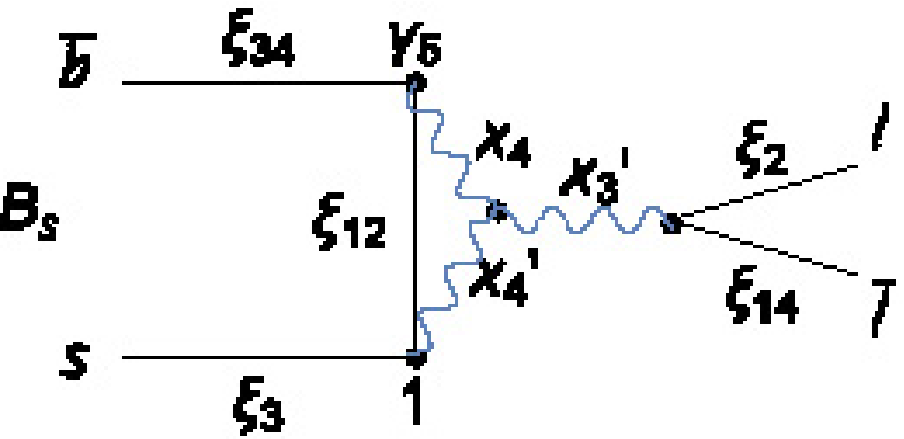}%
\end{center}
\end{minipage}
\hfill
\begin{minipage}[b]{0.47\linewidth}
\begin{center}
\includegraphics[width=6cm,angle=0,clip]{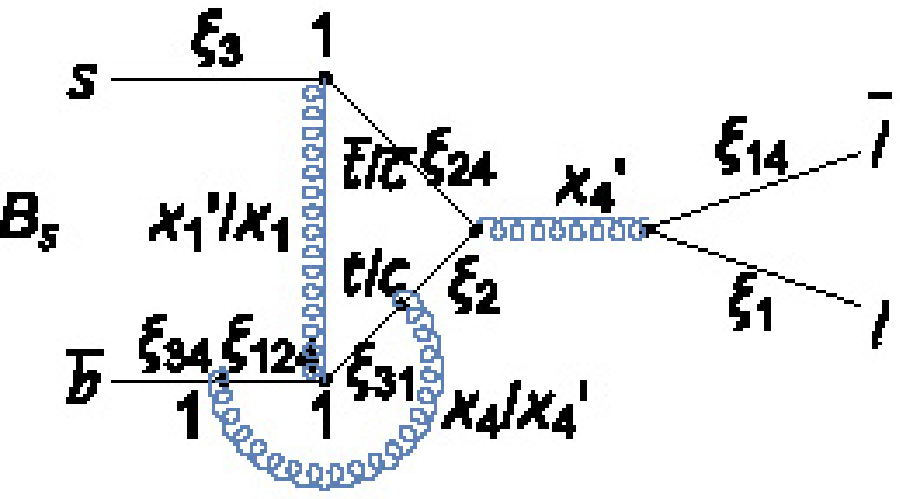}%
\end{center}
\end{minipage}
\begin{minipage}[b]{0.47\linewidth}
\begin{center}
\includegraphics[width=6cm,angle=0,clip]{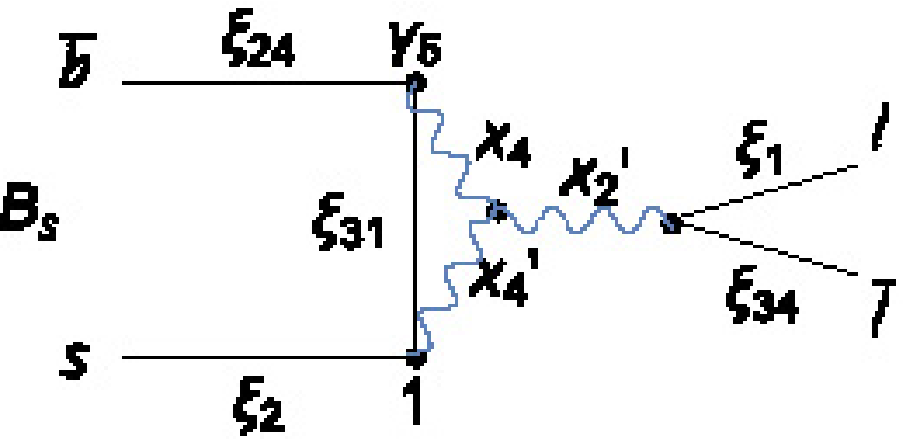}%
\end{center}
\end{minipage}
\hfill
\begin{minipage}[b]{0.47\linewidth}
\begin{center}
\includegraphics[width=6cm,angle=0,clip]{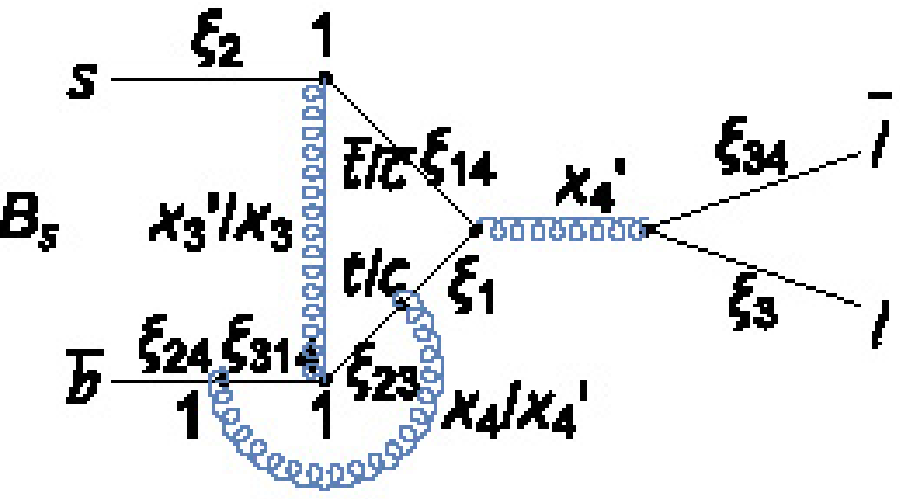}%
\end{center}
\end{minipage}
\begin{minipage}[b]{0.47\linewidth}
\begin{center}
\includegraphics[width=6cm,angle=0,clip]{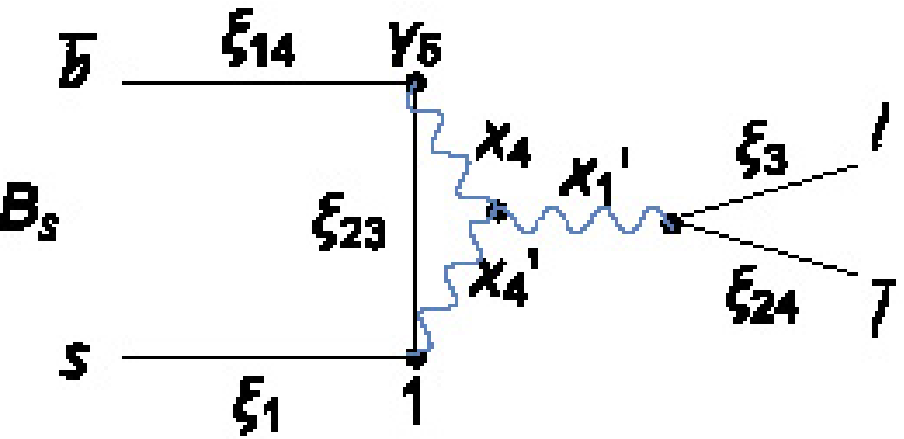}%
\end{center}
\end{minipage}
\hfill
\begin{minipage}[b]{0.47\linewidth}
\begin{center}
\includegraphics[width=6cm,angle=0,clip]{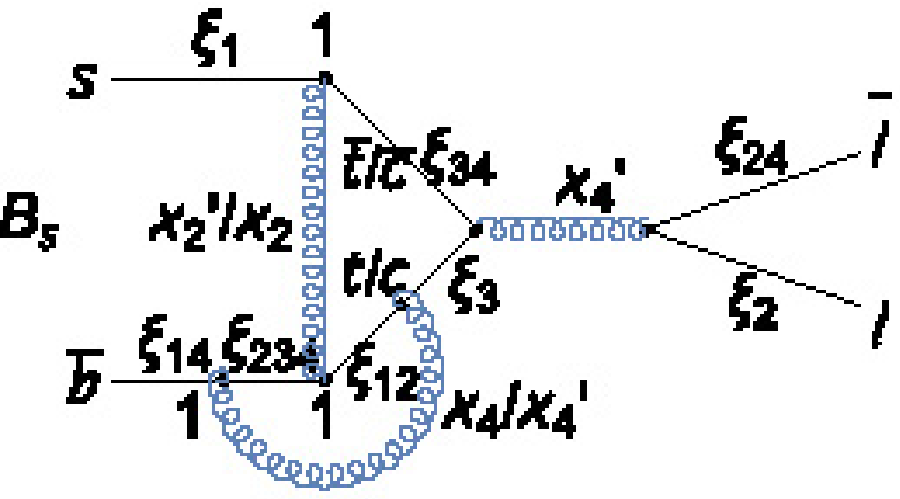}%
\end{center}
\end{minipage}
%
\begin{minipage}[b]{0.47\linewidth}
\begin{center}
\includegraphics[width=6cm,angle=0,clip]{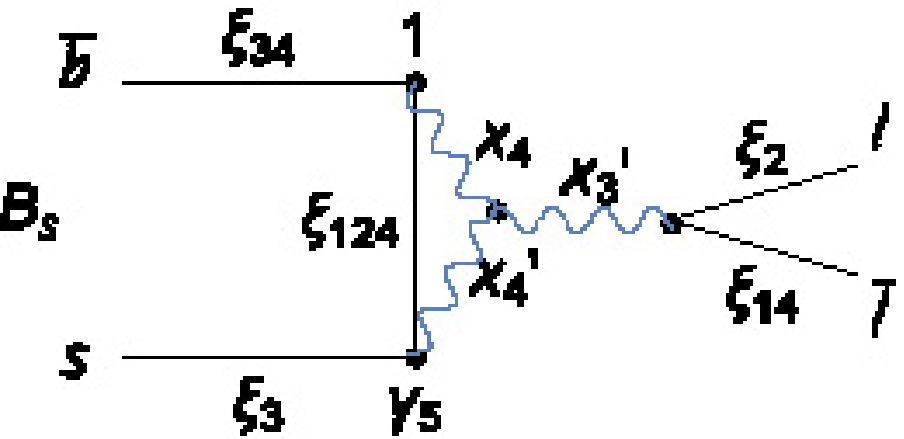}%
\end{center}
\end{minipage}
\hfill
\begin{minipage}[b]{0.47\linewidth}
\begin{center}
\includegraphics[width=6cm,angle=0,clip]{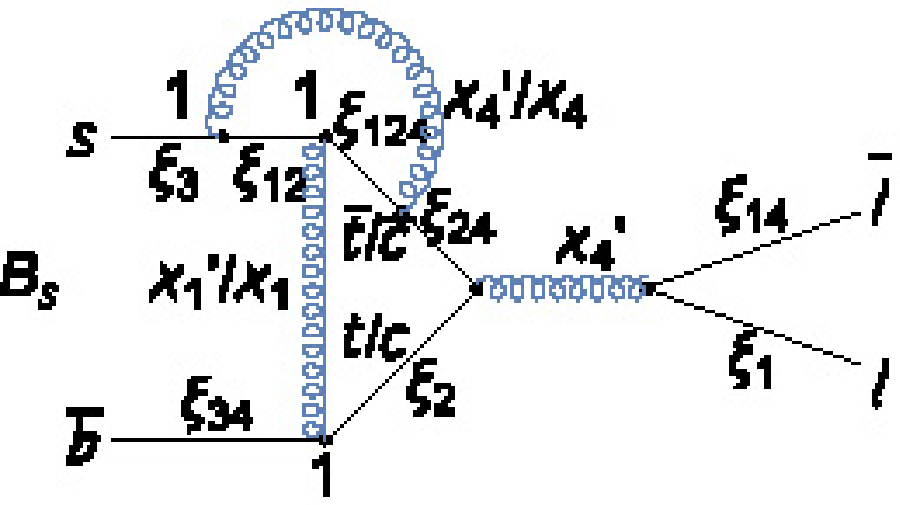}%
\end{center}
\end{minipage}
\begin{minipage}[b]{0.47\linewidth}
\begin{center}
\includegraphics[width=6cm,angle=0,clip]{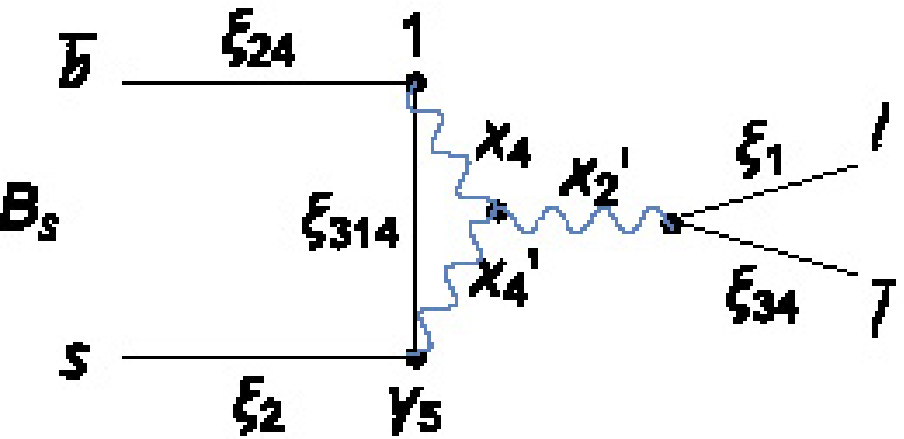}%
\end{center}
\end{minipage}
\hfill
\begin{minipage}[b]{0.47\linewidth}
\begin{center}
\includegraphics[width=6cm,angle=0,clip]{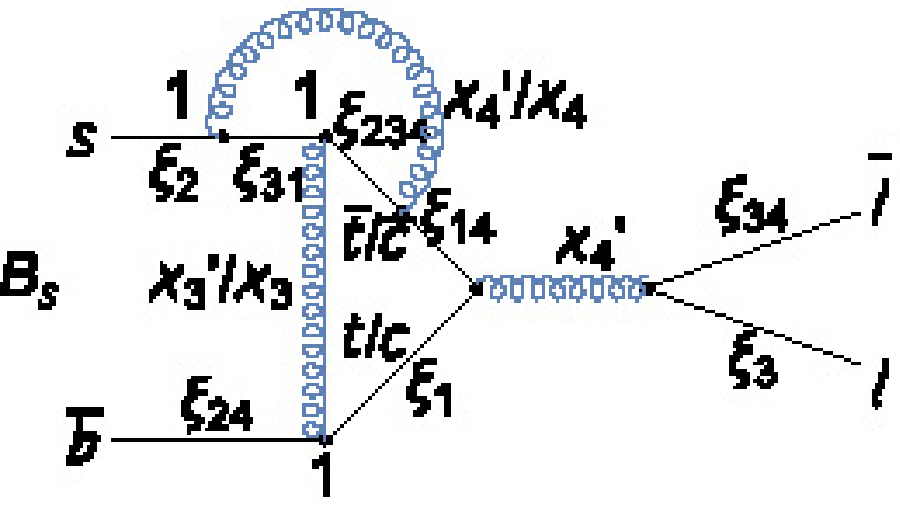}%
\end{center}
\end{minipage}
\begin{minipage}[b]{0.47\linewidth}
\begin{center}
\includegraphics[width=6cm,angle=0,clip]{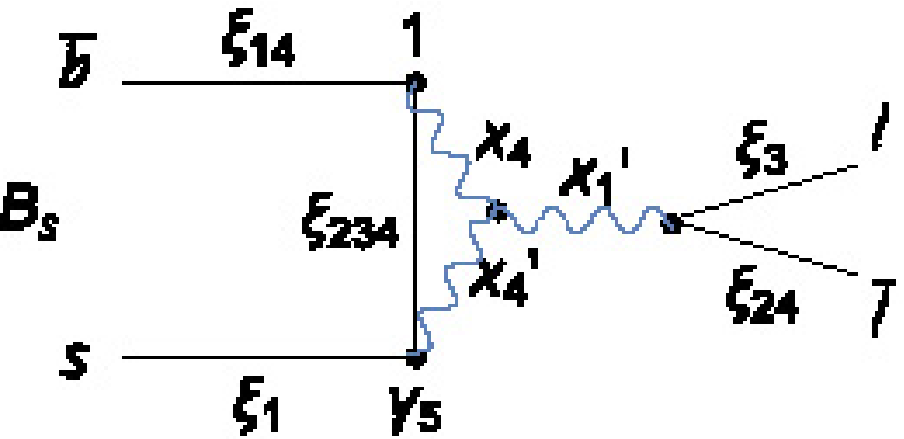}%
\end{center}
\end{minipage}
\hfill
\begin{minipage}[b]{0.47\linewidth}
\begin{center}
\includegraphics[width=6cm,angle=0,clip]{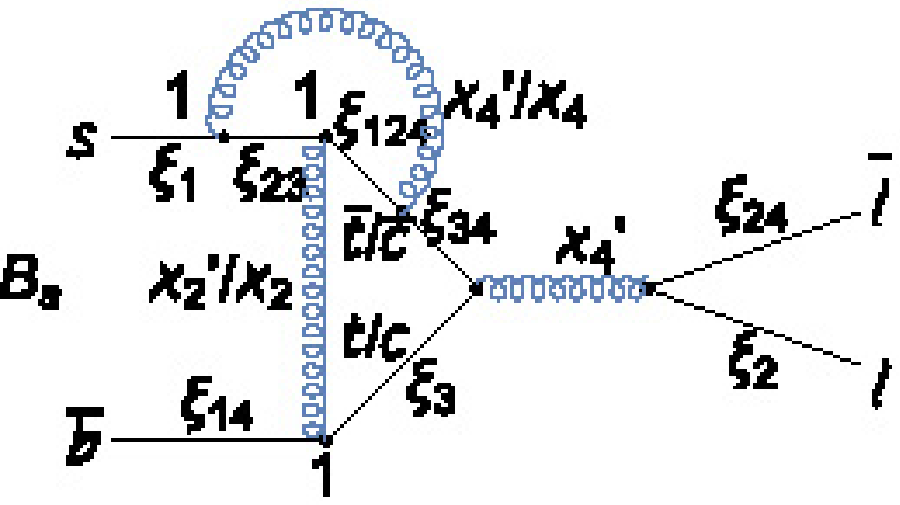}%
\end{center}
\end{minipage}
\caption{$B_s\to \ell\bar\ell$ quark diagrams containing $x_4 {x_4}' {x_k}'$ ($k=1,2,3$) vertices with different $\gamma_5$ positions (left), and $q-\bar q-W$ triangle diagrams with different penguin loop positions (right).} 
\label{bbarsLLa}
\end{figure}

The quark diagrams of $B_d\to\ell\bar\ell$ processes are given in Fig.\ref{bbardLLa} and that of $B_s\to\ell\bar\ell$ processes are given in Fig.\ref{bbarsLLa}.

The qualitative difference of experimental and theoretical branching ratios of $\bar B_{s\mu}$ and $\bar B_{d\mu}$ may originate from the difference of quarks propagating on the triangle of two loop diagrams between $s$ and $\bar b$ and $d$ and $\bar b$.
In the case of $B_s$, since the $s$ quark belongs to the same sector as the $c$ quark, after emitting a vector particle $x_i'$, it can be transformed to an antiquark $\bar c$ and emits a vector particle $x_4'$ which changes to $\ell\bar\ell$, returns to a quark $c$ which emits a vector particle ${x_4}'$. The $c$ quark absorbs the emitted vector particles $x_i'$ and $x_4'$ and transforms to a $\bar b$ quark.

In the case of $B_d$, the $d$ quark is transformed to $\bar t$ quark after emitting a vector particle ${x_i}'$ as shown in the right-hand part of diagrams. The product $V_{ud}V_{ub}^*$ of CKM matrices gives the strength, and the $\bar t$ quark emits an ${x_4}'$ and transforms to a $\bar b$ quark after absorbing the emitted $x_4'$ on the $\gamma_5$ vertex.

The relative intensity of the left-hand side diagrams and the right-hand side diagrams of Figs.\ref{sbardLLa}, \ref{bbardLLa}, and Fig.\ref{bbarsLLa} are not evident. We expect the left-hand side dominates, and reflects the CKM matrix elements, however in $B_s\to \ell\bar\ell$ the right-hand side diagram containing a $c$ quark propagation gives a slight suppression to the left-hand side dominant term, and in $B_d\to \ell\bar\ell$ the right-hand side diagram containing a $t$ quark propagation is expected to give an enhancement to the dominant term.

\section{Discussion and conclusion}
Description of Dirac fermions is not unique. The sub algebra $\Vec R_{3,1}^+$ with a basis 
\[
\{\Vec 1, \Vec e_1 \Vec e_2, \Vec e_1 \Vec e_3, \Vec e_2 \Vec e_3,\Vec e_1\Vec e_4, \Vec e_2 \Vec e_4 ,\Vec e_3 \Vec e_4,\Vec e_1 \Vec e_2 \Vec e_3 \Vec e_4 \}
\] 
where $\Vec e_i$ satisfy $\Vec e_1^2=\Vec e_2^2=\Vec e_3^2=\Vec 1, \Vec e_4^2=-\Vec 1$ is equivalent to Pauli algebra, defined by 
$\Vec e_1\Vec e_4=\sigma_1, {\Vec e_2\Vec e_4}=\sigma_2, {\Vec e_3\Vec e_4}=\sigma_3
$
and
\[
\Vec e_1 \Vec e_2 \Vec e_3 \Vec e_4=\sigma_1\sigma_2\sigma_3=\Vec i.
\]
CKM matrices were derived based on this algebra\cite{Aragon08} %

In the $n$ dimensional linear space $V$ over a field $\Vec F$ and exterior algebra $\wedge V$, octonion appears by defining $\Vec e_1\Vec e_2\Vec e_3=\Vec \ell$, and choosing 
\[
\{\Vec 1,\Vec e_1,\Vec e_2,\Vec e_3, -\Vec e_3\Vec\ell, -\Vec e_2\Vec\ell, -\Vec e_1\Vec\ell,\Vec \ell\}
\]
as the basis of the field\cite{DM98a,DM98b}. The commutation relations of $\Vec e_i$ are not same as those of Cartan's $\xi_i$.

The rules of multiplication of $\Vec e_i$ follow Clifford algebra\cite{Porteous95}. Lounesto\cite{Lounesto01} pointed out that a product of Clifford numbers $\Vec x$ and $\Vec y$ in $n$ dimensional linear space $V$ over a field $\Vec F$ defined by Chevalley\cite{Chevalley65} is 
\[
\Vec x \Vec y=\Vec x\wedge\Vec y+\Vec x\lrcorner \Vec y=\Vec x\wedge\Vec y+B(\Vec x,\Vec y)
\] 
where $\Vec x\wedge \Vec y$ is the antisymmetric product, and $\Vec x\lrcorner\Vec y$ is the contraction which depends on $\Vec F$.

When $\Vec F=\{0,1\}$ and an exterior algebra $\wedge V$ has the basis $\{ 1,\Vec e_1,\Vec e_2,\Vec e_1\wedge\Vec e_2\}$, there are two bilinear forms $B_1(\Vec x,\Vec y)=x_1 y_2$ and $B_2(\Vec x,\Vec y)=x_2y_1$ which do not have a canonical multiplication table.

The time(space) component of $\psi=\xi_4(\xi_i)$ and that of $\phi=\xi_0(\xi_{i4})$ satisfy the commutation relation
\[
\frac{1}{2}(\xi_0\xi_4-\xi_4\xi_0)=1,\quad \frac{1}{2}(\xi_{i4}\xi_i-\xi_i\xi_{i4})=1\quad (i=1,2,3) 
\] 
Similarly, those of $\mathcal C\psi=\xi_{123}(\xi_{ij4})$ and $\mathcal C\phi=\xi_{ij3}(\xi_{ij})$ satisfy the commutation relation 
\[
\frac{1}{2}(\xi_{1234}\xi_{123}-\xi_{123}\xi_{1234})=1,\quad \frac{1}{2}(\xi_{ij}\xi_{ij4}-\xi_{ij4}\xi_{ij})=1 \quad(ij=12,23,31).
\]
Due to this difference of commutation relation of the space components and time components of $\psi$ and $\phi$, we can define unique octonion in $\bf R^8$ space.

We applied Cartan's supersymmetry to analyze the decay branching ratios of $K$ and $B$ mesons to $\pi\ell\bar\ell$ and $\ell\bar\ell$.
In the $B_{s,d}\to \mu^+\mu^-$ process, we assigned vector particles $x_i$ to $Z$ boson of polarization $x_i$ and by allowing transition of $\mathcal C\phi$ and $\mathcal C\psi$ during the propagation of $b\to t\to \bar s$, we could make $B_{s,d}$ composed of large components $\psi$ or $\phi$. 

The suppression of the $B_s\to \mu^+\mu^-$ can be explained, if one takes into account that the $c$ quark and $s$ quark are in the same sector, and correction to the CKM mechanism is large in $B_s\to \mu^+\mu^-$ than in $B_d\to \mu^+\mu^-$. When $s$ or $c$ quarks are not present in the meson system, the vertex of vector particles $x_4-{x_4}'-{x_j}'$($j=1,2,3$) and the triangular loop of top quarks and one vector particle $t-\bar t-{x_j}'$ defines the decay branching ratios.
When $s$ or $c$ quarks are in the meson, the loop of $c-\bar c-{x_j}'$ modifies the $t-\bar t-{x_j}'$ contribution.

In the case of $B_d\to \mu^+\mu^-$, since $d$ quark is not in the sector of $c$, only the loop of $t-\bar t-{x_j}'$ contribute, and if it adds the contribution of the one loop of $x_4-{x_4}'-{x_j}'$($j=1,2,3$), we can explain the enhancement of $B_d\to \mu^+\mu^-$. 
 
In the processes of $B_s\to \mu^+\mu^-$ via exchange of Majorana neutrino\cite{CMS14}, one allows a transition from $\nu$ to $\bar\nu$.
In our model, $|e^-,\nu_e)_L , |\mu^-,\nu_\mu)_L, |\tau^-,\nu_\tau)_L$ and their charge conjugates make sectors, and a neutrino in the sector of $|\mu^-,\nu_\mu)_L$ can transform to $|\mu^+,\bar\nu_\mu)^*$ or  $|\mu^+, \bar\nu_\mu)^{**}$ in the notation of \cite{SF15a}, which can be interpreted as $\bar\nu_\mu$ in different triality sectors, which would not be detected by vector particles or leptons in the sector of the initial $|\mu^-, \nu_\mu)_L$, since for detection of a particle $a_i^{k(j)}$ in the sector $k$, corresponding necessary condition $N_j$ also need to be transferred \cite{Bitbol88} from different sectors, and we may be allowed to ignore the $\nu\to\bar\nu$ transition.

If our electromagnetic detector can detect electromagnetic field transformed by $G_{23}$ but cannot detect electromagnetic field transformed by other four transformations, and there are six lepton sectors 
\[
|e,\nu_e)^*,|\mu,\nu_\mu)^*,|\tau,\nu_\tau)^*, |e,\nu_e)^{**},|\mu,\nu_\mu)^{**},|\tau,\nu_\tau)^{**}
\]
which cannot be detected by our detectors, the presence of dark matter can be explained.


 We observed that starting from the meson wave function given by the large component of a quark and the large component of an anti-quark, qualitative features of the strength of the decay branching ratios can be obtained from Cartan's supersymmetry.

Cartan's octonion satisfies a triality automorphism that is supersymmetric, which octonions of \cite{DM98a,DM98b}  dont satisfy. In our system of fermions and vector fields with interaction ${^t\phi}C(1-\gamma_5)X\psi$, the operator $\gamma_5$ yields coupling between a large component of a quark and a large component of an antiquark which enhances certain $B$ meson decay processes. It makes the branching ratio of $B$ meson decays different from the CKM model.

 If in the univese there are world which are transformed by $G_{ij}$ and $G_{ijk}$, and our electromagnetic detector can detect the world transformed by $G_{12},G_{13},G_{123}$ and $G_{132}$, and the uncertainty principle applys not only in our world but also whole universe, we can understand the presence of dark matter.

Departure from the standard model of $B_s\to\mu\bar\mu$ and $B_d\to\mu\bar\mu$ is due to the large correction from the $c$ quark in the decay of $B_s$ meson. We ignored $K\to\pi\nu\bar\nu$ and $B\to D\ell\bar\nu$ etc, whose information will serve for establishing the model of $B$ mesons based on Cartan's supersymmetry.

\vskip 1 true cm

\end{document}